\documentclass[aps,prd,amssymb,showpacs,floatfix,twocolumn,superscriptaddress,nofootinbib]{revtex4-1}
\usepackage{amsmath,amsfonts,graphics,epsfig,amssymb}

\begin{document}
\title{%
Transition from a relativistic constituent-quark model to the
quantum-chromodynamical asymptotics: a quantitative description of the pion
electromagnetic form factor at intermediate values of the momentum
transfer}

\author{S.V.~Troitsky}

\affiliation{Institute for Nuclear
Research of the Russian Academy of Sciences, 60th October Anniversary
Prospect 7a, Moscow 117312, Russia}

\email{st@ms2.inr.ac.ru}

\author{V.E.~Troitsky}

\affiliation{D.V.~Skobeltsyn Institute of Nuclear Physics,
M.V.~Lomonosov Moscow State University, Moscow 119991, Russia}

\email{troitsky@theory.sinp.msu.ru}


\date{v.2: October 30, 2013}

\begin{center}
\begin{abstract}
We adopt a non-perturbative relativistic constituent-quark
model for the $\pi$-meson electromagnetic form factor, which
have successfully predicted experimental results, and supplement it
with the effective momentum-dependent quark mass to study quantitatively
the transition to the perturbative QCD asymptotics.
The required asymptotical behaviour (including both the $Q^{-2}$ fall-off
and the correct coefficient) settles down automatically when the quark mass
is switched off; however, the present experimental data on the form factor
suggest that this cannot happen at the values of the momentum transfer
below $\sim 10$~GeV$^{2}$. The effective constituent-quark mass below this
scale acquires substantial non-perturbative contributions.
\end{abstract}
\end{center}
\maketitle


\section{Introduction}
\label{sec:intro}

Bound states of light quarks, and the $\pi$ meson in particular, represent
a challenging testbed for our understanding of the strong interaction.
Theoretical approaches to their description are split into two directions.
From the high-energy side, the quantum chromo\-dynamics (QCD), which is
widely believed to be a fundamental theory of the strong force, becomes
strongly coupled at the relevant energy scales, so trustable perturbative
calculations help a little in quantitative description of precise
low-energy data, which there is no lack of. From the low-energy side, a
number of successful models to describe the data have been developed. To
be quantitative, they necessarily require some phenomenological input.
None of these models can be consistently and quantitatively derived from
the QCD lagrangian, therefore a gap between the two approaches emerges.
The purpose of the present paper is to contribute to filling this gap by
making a bridge between a successful low-energy model and the high-energy
QCD calculation. We choose the electromagnetic form factor of the charged
pion as the observable to study.

While most of the approaches fall into one of two groups, that is
either low-energy (soft) or  high-energy (hard) ones, and therefore
cannot address the intermediate (transition) range of typical energies or
momenta, there are some remarkable attempts to cover all energy scales
within a single framework. One is the numerical non-perturbative QCD
realized at the lattice. Despite a considerable success, technical
difficulties presently prevent it from obtaining reliable quantitative
results for light mesons at intermediate momenta. Another approach is
based on the holographic duality between strongly and weakly coupled
theories which, for QCD, has not been rigorously proven, though works well
in a number of phenomenological applications. By definition, any duality
construction is uneasy to implement quantitatively at intermediate
energies, where both dual theories are moderately strongly coupled.

Our goal here is less ambitious. We adopt frameworks of a particular
relativistic constituent-quark model with low-energy phenomenological
parameters included and study how the QCD asymptotics is reached within
this particular model.

The asymptotics of the pion electromagnetic form factor $F_{\pi}$ at
momenta transfer $Q^{2}\to \infty$ has been determined
\cite{FarrarJackson, EfremovRadyushkinAs, LepageBrodskyAs}, in the QCD
frameworks, as
\begin{equation}
Q^{2} F_{\pi}(Q^{2}) \to 8\pi \alpha_{\rm s}^{\rm 1-loop}(Q^{2})
f_{\pi}^{2},
\label{Eq:as}
\end{equation}
where $\alpha_{\rm s}^{\rm 1-loop}(Q^{2})=4\pi/\left(\beta_{0}
\log\left(Q^{2}/\Lambda_{\rm QCD}^{2} \right) \right)$
is the one-loop running strong coupling constant,
$\beta_{0}=11-2N_{f}/3$
is the first beta-function coefficient, $N_{f}$ is the number of
active quark flavours and $f_{\pi}\approx 130$~MeV
\cite{PDG} is the pion decay constant. It is important to note that this
asymptotical behaviour, consistent with the quark counting rules
\cite{Qcounting1, Qcounting2}, includes the one-loop coupling only and is
to be modified whenever the one-loop approximation fails, but not by means
of a simple replacing of $\alpha_{\rm s}$ with its more precise value.
Involved QCD calculations have been performed to obtain corrections to
Eq.~(\ref{Eq:as}), see e.g.\ \cite{Bakulev?}. The QCD does not predict the
value of $Q^{2}$ at which this asymptotics should be reached.

On the other hand, the experimental data on $F_{\pi}$ (see e.g.\
Ref.~\cite{exp-data} for a review) are well described by a number of
low-energy nonperturbative models, provided some phenomenological
parameters are tuned. In most cases, these models do not attempt to
describe the soft-hard transitions (see however a few important exceptions
discussed in Sec.~\ref{sec:formfactor:results}). To our best knowledge,
none of the successful low-energy models have described quantitatively and
without dedicated tuning of parameters how the asymptotics (\ref{Eq:as})
settles down, given the experimental data. We attempt to do it in the
present work.

We start with a well-established Poincar\'e-invariant constituent-quark
model ~\cite{KrTr-EurPhysJ2001, KrTr-PRC2002, KrTr-PRC2003, KrTr-PRC2009,
EChAYa2009} which has, for our purposes, the following three advantages:
\begin{itemize}
 \item[(i)~]\textbf{predictivity}:
starting from the experimental data on
$F_{\pi} (Q^2)$
at $Q^2 \lesssim 0.26$~GeV$^2$
\cite{Amendolia}, this approach allowed to predict,
in  1998, the values of the pion form factor for
the extended range of higher momentum transfer \cite{KrTr-EurPhysJ2001}.
The experimental data obtained later (see
Ref.~\cite{exp-data} and
references therein) for the range of $Q^2$ larger by an order of magnitude
coincide precisely with the prediction of Ref.~\cite{KrTr-EurPhysJ2001}
without any further tuning of parameters;
\item[(ii)~]\textbf{robustness}: the behaviour of $F_{\pi}(Q^{2})$ at
moderate and high $Q^{2}$ does not depend on the selected wave function
and is determined by the constituent-quark mass
only~\cite{KrTr-EurPhysJ2001};
\item[(iii)~]\textbf{asymptotical behaviour}: it has been mathematically
proven~\cite{KrTr-asymp} that in the limit $Q^{2}\to \infty$ and the
constituent-quark mass $M\to 0$, the QCD asymptotical behaviour,
$Q^{2}F_{\pi}(Q^{2})\sim\mbox{const}$, is obtained in this model.
\end{itemize}
The points (ii), (iii) suggest a following
concept \cite{KrTr-PRC2009, Kiss} to study the soft-hard transition for
$F_{\pi}$. One should adopt
the successful low-energy framework and supplement it with a
$Q^{2}$-dependent quark mass $M$, which should be equal to a
constituent-quark mass at $Q^{2}\to 0$ but should fall at $Q^{2}\to
\infty$ to provide the correct $M\to 0$ limit. In this work, we accept a
dependence of $M(Q^{2})$ motivated by other studies and fulfill this
program.

As a result, we obtain a quantitative description of the hard-soft
transition. We will see that, given the experimental data, this transition
should not take place at low $Q^{2}$. Therefore, the data on
$F_{\pi}$ constrain possible models for $M(Q^{2})$ and indicate the
presence of large nonperturbative contributions to the light quark mass at
least up to $Q^{2}\sim 7$~GeV$^{2}$.

The rest of the paper is organized as follows. In Sec.~\ref{sec:model}, we
formulate and briefly discuss the model we use. In particular, the concept
of the relativistic constituent-quark model is given in
Sec.~\ref{sec:model:constituent}, together with references to more
detailed descriptions. Sec.~\ref{sec:model:mass} discusses the selected
model for $M(Q^{2})$ and puts it in the context of other approaches used
in the literature. We proceed with the calculation of the pion form factor
in Sec.~\ref{sec:formfactor}, where we first discuss
(Sec.~\ref{sec:formfactor:param}) parameters of the model, their
fixing/constraining, and the remaining freedom. Results for $F_{\pi}$,
obtained within these constraints, are presented and discussed in
Sec.~\ref{sec:formfactor:results}, where the comparison with other
approaches is also given. We briefly conclude in Sec.~\ref{sec:concl} and
list some cumbersome formulae in the Appendix.

\section{The model}
\label{sec:model}

\subsection{The relativistic constituent-quark model}
\label{sec:model:constituent}
 Our method
is a version of the instant form of the
Poincar\'e invariant constituent-quark model (PICQM).
This model is described in detail in
Refs.~\cite{KrTr-EurPhysJ2001, KrTr-PRC2002, KrTr-PRC2003, KrTr-PRC2009,
EChAYa2009}. Briefly, the model is constructed as follows. In the instant
form (IF) of the Relativistic Hamiltonian Dynamics (see e.g.\ Ref.
\cite{KeisterPolyzou}) one considers two non-interacting one-particle
states. Then one separates the center-of-mass motion of the system as a
whole and, by means of the Wigner-Eckart theorem for the Poincar\'e group,
extracts the reduced matrix elements, that is form factors
\cite{TMF2005}. To include the interaction, one finds solutions to the
Muskhelishvili-Omnes type equations. These solutions represent wave
functions of constituent quarks.

It is important to notice that the approach we use differs from the
IF \textit{per se} but it was rather fruitfully complemented by
the so-called Modified Impulse Approximation (MIA), see
Ref.~\cite{KrTr-PRC2002}. MIA is constructed by making use of a
dispersion-relation approach and removes certain (often quoted)
disadvantages of the IF. In particular, our framework is fully
relativistic.

The final result for the pion form
factor,
$F_{\pi}(Q^{2})$, is given by a double integral representation,
\begin{equation}
F_\pi(Q^2)=\int\!
\mathrm{d}\sqrt{s}\,
\mathrm{d}\sqrt{s'}\,
\varphi(k)\,g_0(s,Q^2,s')\,\varphi(k')\;,
\label{Eq:Fpi-integral}
\end{equation}
where $s=4(k^2+M^2)$, $\varphi(k)$ is the PICQM pion wave function and
$g_0(s,Q^2,s')$ is the free two-particle form factor. The latter may be
obtained explicitly by the methods of relativistic kinematics and is a
relativistic invariant function; its actual form is rather cumbersome and
is given in Appendix for reference.

In Eq.~(\ref{Eq:Fpi-integral}),
$$
\varphi(k) = \sqrt[4]{s}\,u(k)k
$$
and a phenomenological wave function
$u(k)$ is to be supplied. Fortunately, as it has been shown in
Ref.~\cite{KrTr-EurPhysJ2001}, the behaviour of the form factor is
insensitive to the choice of the wave function provided a phenomenological
boundary condition (correct pion charge radius $\langle
r_{\pi}^{2}\rangle^{1/2}$) is satisfied and the correct value of the pion
decay constant $f_{\pi}$ is reproduced. In the present work, we use the
power-law type wave function (see e.g.\ Ref.~\cite{wave-function}),
$$
u(k)=N \left(k^2 /b^2 +1  \right)^{-3},
$$
where the normalization coefficient $N=16\sqrt{2}/\sqrt{7\pi b^{3}}$ and
$b$ is a phenomenological parameter related to the confinement scale.

Several other wave functions were shown in Refs.~\cite{KrTr-EurPhysJ2001,
KrTr-PRC2009} to give the same result for the form factor, provided its
low-energy behaviour is fixed by fitting the pion charge radius with a
single wave-function parameter, $b$ in our case. The behaviour of
$F_{\pi}(Q^{2})$ at $Q^{2}\gtrsim 1$~GeV$^{2}$, given this fit, is
completely determined by $M$, as it has been demonstrated in
Ref.~\cite{KrTr-EurPhysJ2001} where calculations have been performed for
various $M$ and different wave-function parametrizations.

The asymptotical behaviour of  $F_{\pi}(Q^{2})$ has been studied in
Ref.~\cite{KrTr-asymp} by considering, mathematically, the $Q^{2}\to
\infty$ limit. In this limit, the original model has two
essential dimensionful parameters, the quark mass $M$ and the confinement
scale $b$. It has been demonstrated that at $M\to 0$, that is $M\ll b$, the
QCD asymptotical quark-counting rules are reproduced, $Q^{2}
F_{\pi}(Q^{2}) \to \mbox{const}$. As we will demonstrate below in the
present work (Sec.~\ref{sec:formfactor:results}) by numerical
calculations, the constant of Eq.~(\ref{Eq:as}) is also properly
reproduced in the $M\to 0$ limit. This is a remarkable success of the
model.

\subsection{Model for the effective $M(Q^{2})$}
\label{sec:model:mass}

To merge the correct low-energy behaviour (which suggests the
constituent-quark mass $M$ of order the confinement scale $b$) with the
correct QCD asymptotics (which requires $M\ll b$ as we have just discussed
above), one should describe the transition between the two regimes by
means of a momentum-dependent effective quark mass $M(Q^{2})$. We will now
discuss possible approaches to determination of this function and
formulate our prescriptions.

Since the current-quark mass $m$, a parameter of the QCD lagrangian,
experiences the renormalization-group running and is thus scale-dependent,
one might be tempted to identify this $m(Q^{2})$ dependence with the one
we are seeking for $M$. If this were true, the mass $m$, when running from
large values of $Q^{2}$ down to $Q^{2}\sim \Lambda^{2}_{\rm QCD}$, where
the QCD coupling becomes really strong, should acquire large
non-perturbative corrections which in turn should increase the mass by two
orders of magnitude. The renormalization-group equation for $m$ has the
schematic form
\[
\frac{dm^{2}}{d\log Q^{2}}=m^{2} \gamma(\alpha_{\rm s}),
\]
where $\gamma(\alpha_{\rm s})$ does not depend explicitly on $m$. For a
small $m$ this indicates a small derivative, so light particles should
remain light unless extreme nonperturbative effects blow up
$\gamma(\alpha_{\rm s})$. The latter possibility is in principle possible,
and some authors attempted to run $m(Q^{2})$ all the way down to $Q^{2}=0$
under certain assumptions about non-perturbative renormalization (see e.g.\
Ref.~\cite{Step}). The mass $m(0)$ obtained in this way was significantly
lower than the constituent-quark mass $M$ inferred from phenomenological
models. This is not surprising, however, because, as is most clearly
stated in the ``Quark masses'' review \cite{PDG-qmass} of the Particle
Data Group, the consituent-quark masses \textit{``make sense in the
limited context of a particular quark model, and cannot be related to the
quark mass parameters of the Standard Model''}. Therefore, we should
consider the $Q^{2}$ dependence  of the constituent-quark mass $M$
starting from its infrared value and cannot impose the QCD value of $m$ as
a precise boundary condition for $M(Q^{2})$ defined in a particular model.
The function $M(Q^{2})$ should decrease at large $Q^{2}$ to reflect fading
of non-perturbative effects, but, being defined in a model-dependent way,
it should not be identified with $m$ anywhere.

There are some remarkable approaches to quantifying this non-perturbative
mass ``running''. One is based on the gap equation supplemented by the
effective confining propagator for gluons and a one-gluon exchange for
constituent quarks \cite{Cornwall, Posled}. The resulting $M(Q^{2})$
function was obtained numerically in Ref.~\cite{Posled} by means of a
solution to a certain integral equation. It depends on the following four
parameters: the confinement scale, two infrared dynamical mass
scales and $\Lambda_{\rm QCD}$. While all these
parameters are of the same order, one should vary all of them to obtain
phenomenologically acceptable results.

Another approach to $M(Q^{2})$ is based on the Dyson-Schwinger equations
in QCD \cite{RobertsWilliams, TandyRev}. It requires assumptions for the
behaviour of the strong coupling constant in the non-perturbative domain
which might be quantified with introducing several free parameters (cf.\
five parameters
described in Ref.~\cite{nucl-th/0005015}).

Ref.~\cite{Kiss} explored varying quark mass $M(Q^{2})$ in the context of
a light-front constituent-quark model, the approach close to the one we
follow here. They adopted a parametrization
\begin{equation}
M(Q^{2})=M_{\rm uv}+
\left(M_{\rm ir} - M_{\rm uv} \right)
\frac{1+{\rm e}^{-\mu^{2}/\lambda^{2}}}{1+{\rm
e}^{(Q^{2}-\mu^{2})/\lambda^{2}}}
L(Q^{2})
\label{Eq:mass-run}
\end{equation}
with $L(Q^{2})\equiv 1$, which describes a step-like transition from
$M(0)=M_{\rm ir}$ to $M(\infty)=M_{\rm uv}$ at $Q^{2}\sim \mu^{2}$, while
the parameter $\lambda$ is responsible for the smoothness of the
transition ($\lambda=0$ corresponds to a step). Again, this functional
form has four parameters. This form has been subsequently used in other
studies \cite{hep-ph/0608148}.

\begin{figure}
\centering
\includegraphics[width=0.8\columnwidth]{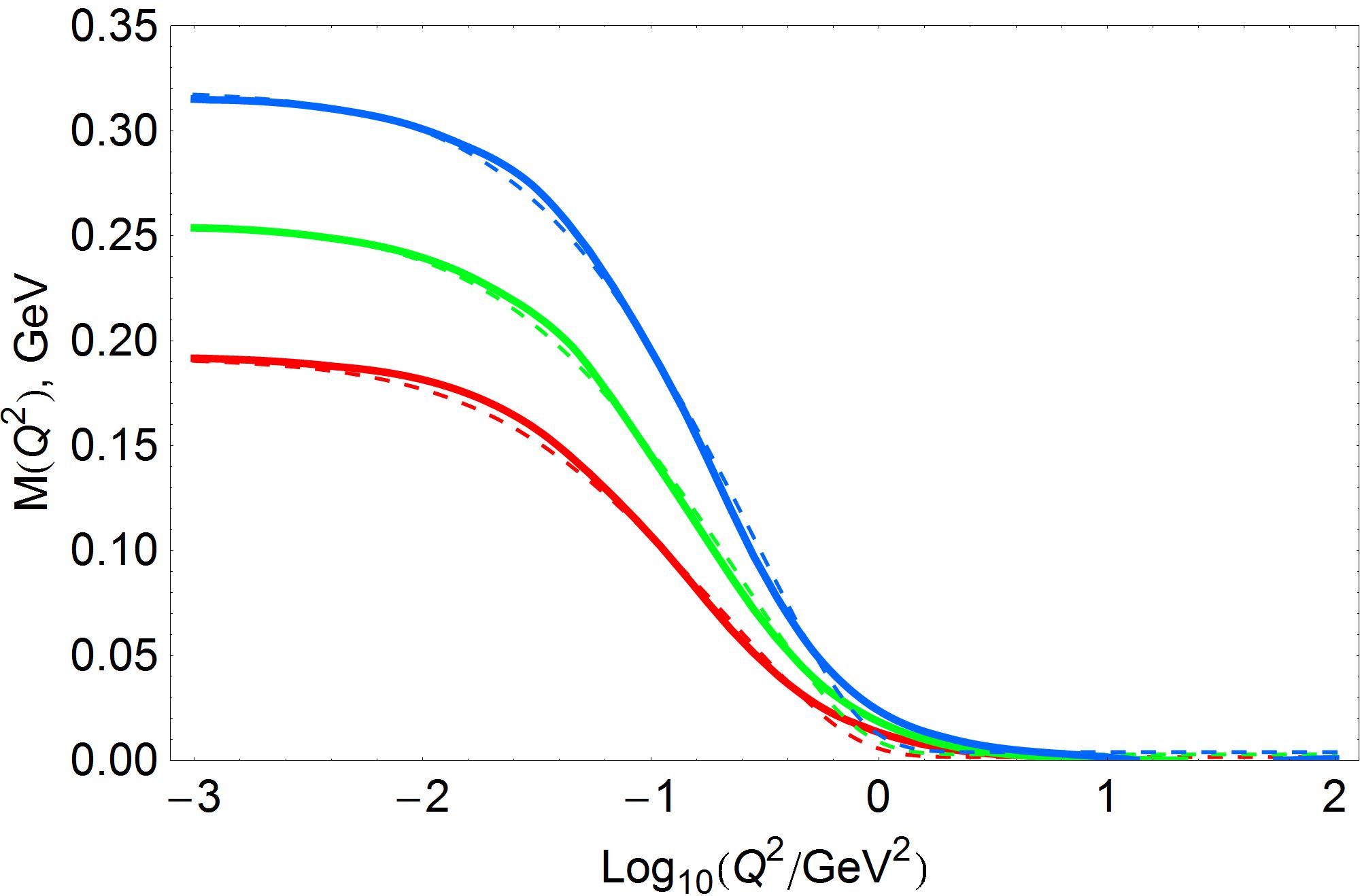}
\caption{Comparison of our approximations for $M(Q^{2})$,
Eqs.(\ref{Eq:mass-run}), (\ref{Eq:mass-run-log}) (dashed lines), with the
results of the full calculation of Ref.~\cite{Posled} (thick full lines),
for various sets of parameters. }
\label{fig:approx-mass}
\end{figure}

We note that changes in free parameters result in considerable variations
of the $M(Q^{2})$ profile in any of the frameworks, while evaluation of
$M(Q^{2})$ for each set of parameters, which we have to perform multiple
times in our approach, represents a
serious numerical task if we choose a dynamical model like
Refs.~\cite{Posled, nucl-th/0005015}. Therefore we, in a way similar to
Ref.~\cite{Kiss}, also choose an ad hoc
parametrization~(\ref{Eq:mass-run}); however, we modify it by setting
\begin{equation}
L(Q^{2})=\frac{1}{1+\log \frac{Q^2+\mu^2}{\mu^2}}
\label{Eq:mass-run-log}
\end{equation}
to better approximate available examples of $M(Q^{2})$ obtained by
detailed calculations, as illustrated in Fig.~\ref{fig:approx-mass}.
We note that Eq.(\ref{Eq:mass-run-log})  does not introduce additional
free parameters with respect to Eq.(\ref{Eq:mass-run}) but improves
significantly the approximation of results of Ref.~\cite{Posled} by
imitating the one-gluon-exchange logarithmic contribution.

\section{The pion form factor}
\label{sec:formfactor}

\subsection{Parameters and constraints}
\label{sec:formfactor:param}
Altogether, the model we use here have the following parameters: the
wave-function confinement scale $b$; the infrared and ultraviolet mass
asymptotics $M_{\rm ir}$, $M_{\rm uv}$ and the two parameters $\mu$ and
$\lambda$ which describe the shape of $M(Q^{2})$,
Eqs.~(\ref{Eq:mass-run}), (\ref{Eq:mass-run-log}). As we have discussed
above in Sec.~\ref{sec:model:constituent}, the QCD asymptotics requires
$M_{\rm uv}=0$, that is we are left with four
parameters\footnote{Numerical instabilities prevent us from taking $M_{\rm
uv}=0$ precisely in calculations. All numerical results presented below
were obtained for $M_{\rm uv}=3.5\times 10^{-6}$~GeV. The two additional
parameters, $s_{q}$ and $c$, which describe the departure from the
point-quark approximation, cf.\ Appendix, are taken from the previous
studies~\cite{KrTr-PRC2002, KrTr-PRC2009} and kept fixed.}. Two of them,
$M_{\rm ir}$ and $b$, correspond to parameters we had in the model with
constant $M$; like previously, we tune them to reproduce two observable
quantities: the pion decay constant $f_{\pi}$ and its charge radius
$\langle r_{\pi}^{2} \rangle^{1/2}$.

The expression for $f_{\pi}$ obtained in Ref.~\cite{Jaus},
\begin{equation}
f_\pi = \frac{M\,\sqrt{3}}{\pi}\,\int\,\frac{k^2\,dk}{(k^2 +
M^2)^{3/4}}\, u(k) ,
\label{Eq:fpi}
\end{equation}
was shown~\cite{KrTr-PRC2002} to be valid in our model. The derivation of
Eq.~(\ref{Eq:fpi}) implies the zero-momentum limit~\cite{KrTr-PRC2002},
therefore one should identify $M=M_{\rm ir}$ in this formula. We use the
most recent Particle Data Group~\cite{PDG} value of $f_{\pi}=130.41\pm
0.20$~MeV.

The pion charge radius is related to the form factor as
\begin{equation}
\langle r_{\pi}^{2} \rangle^{1/2}
=-6 \left.
\frac{dF_{\pi}(Q^{2})}{d Q^{2}}
\right|_{Q^{2}=0}.
\label{Eq:MSR}
\end{equation}
Hereafter, we use the experimental data on $F_{\pi}$ discussed in
Ref.~\cite{exp-data}. In particular, we exclude some of the early
(1976-1978) measurements which, according to Ref.~\cite{exp-data},
may have large and unknown systematic uncertainties. Since the Particle
Data Group world average value of $ \langle r_{\pi}^{2} \rangle^{1/2} $ is
strongly affected by these old data, we take instead the most precise
value of Ref.~\cite{Amendolia}, $ \langle r_{\pi}^{2} \rangle^{1/2} =0.663
\pm 0.006 $~fm. This is consistent with using the low-momentum data points
of Ref.~\cite{Amendolia} when constraining $\mu$ and $\lambda$ as
discussed
below.

Equations (\ref{Eq:fpi}) and (\ref{Eq:MSR}), both related to the $Q^{2}\to
0$ limit, allow us to fix $b=0.6$~GeV  and $M_{\rm ir}=0.22$~GeV. Not
surprisingly, these values are very close to those used in previous works
on the same model ($M_{\rm ir}$ is precisely the same while a minor change
in $b$ reflects the change in the world-average value of $f_{\pi}$ used).
The two parameters $\mu$ and $\lambda$ remain unconstrained at this step.

\begin{figure}
\centering
\includegraphics[width=0.75\columnwidth]{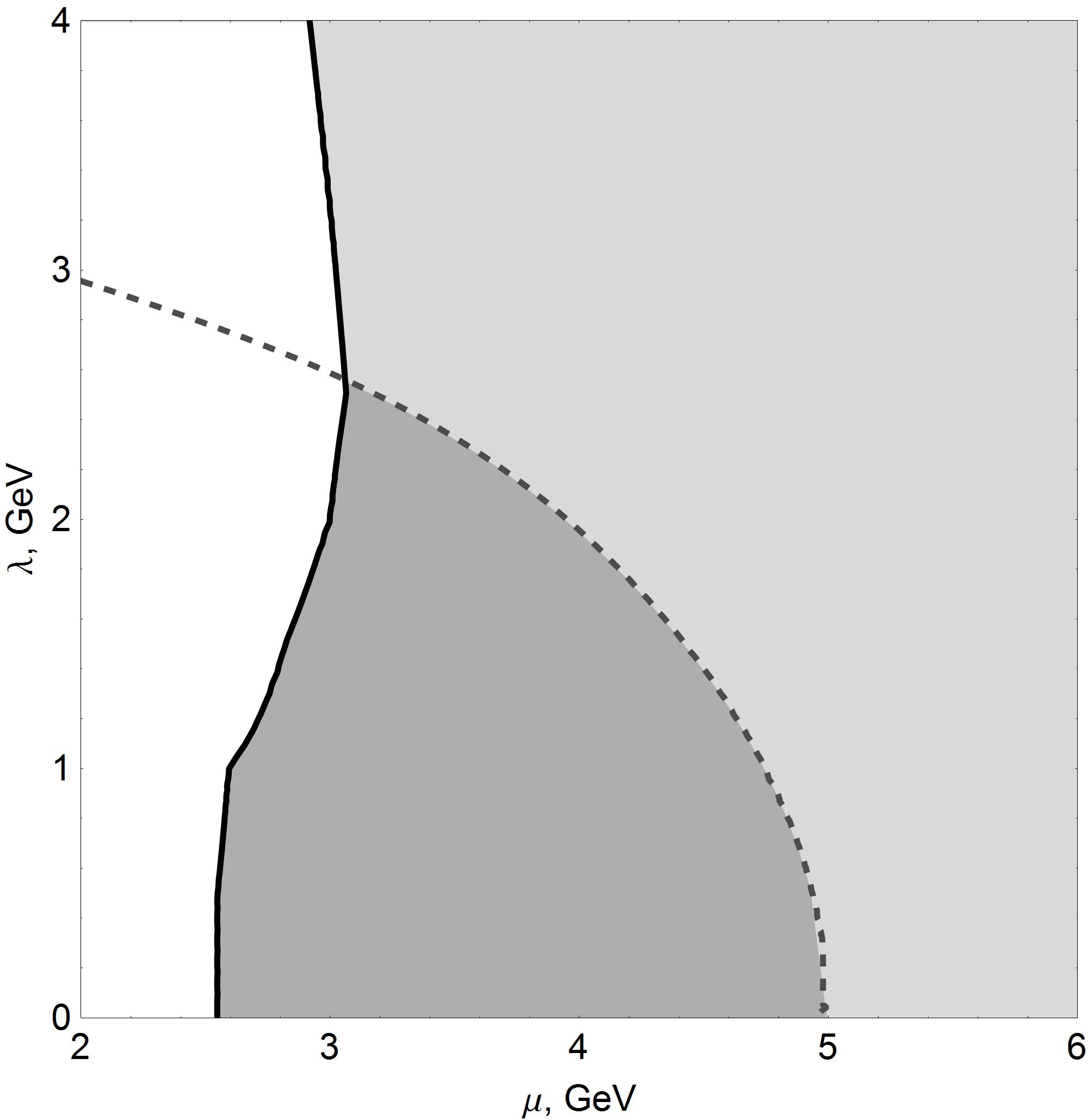}
\caption{Constraints on the parameters $\mu$ and $\lambda$. Thick full line
is the 95\% C.L. limit from the experimental data on $F_{\pi}$; dashed
line is the condition of perturbativity at 25~GeV$^{2}$ (see text).
Parameters in the gray region satisfy the experimental constraint; those
in the dark-gray region satisfy both. }
\label{fig:chi2}
\end{figure}

To proceed further, we note that the experimental values of
$F_{\pi}(Q^{2})$ at all momentum transfers covered by the data, that is
$Q^{2} \le 2.45$~GeV$^{2}$ for the data we use, are in a perfect
agreement~\cite{exp-data, KrTr-PRC2009} with the
predictions~\cite{KrTr-EurPhysJ2001} of the model with $M=\mbox{const}$.
Given the fact that the behaviour of $F_{\pi}(Q^{2})$ is determined by
$M$, see Sec.~\ref{sec:model:constituent}, significant deviations from
$M_{\rm ir}$ below $Q^{2}\sim 2.5$~GeV$^{2}$ may spoil the agreement.
Therefore, we determine a constraint on $\mu$ and $\lambda$ from the
agreement with experimental data on $F_{\pi}$, quantified as the 95\%
confidence-level contour on the $(\mu, \lambda)$ plane obtained by the
standard $\chi^{2}$ analysis (Fig.~\ref{fig:chi2}).
For the determination of this contour, we used the data described in
Ref.~\cite{exp-data}. These data are presented in
Fig.~\ref{fig:Fpi6GeV} and the references are given in its caption.

\begin{figure}
\centering
\includegraphics[width=0.85\columnwidth]{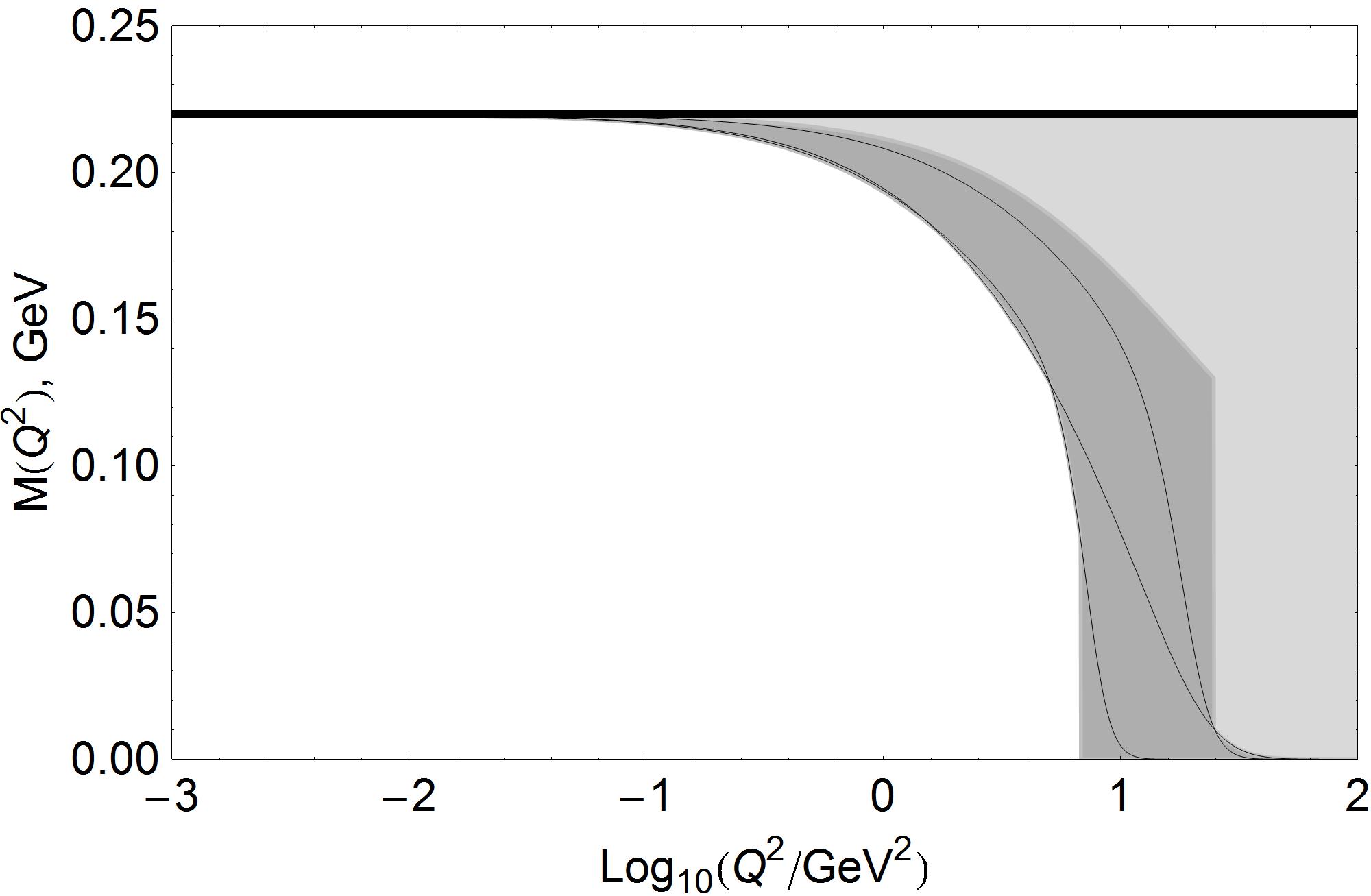}
\caption{
Examples of $M(Q^{2})$ functions (thin lines) and the full
allowed range (gray shadow) determined by the experimental constraints on
$\mu$ and $\lambda$. Imposing additional constraint of perturbativity at
high $Q^{2}$, see text, restricts the allowed range to the dark-gray
region.}
\label{fig:run-mass}
\end{figure}

This however leaves unconstrained the situation when $M(Q^{2})$ remains
large at arbitrary high $Q^{2}$, since the result with $M=M_{\rm
ir}=\mbox{const}$ gives an excellent description of the existing data. One
may think of an additional, optional constraint on $(\mu,
\lambda)$ related to perturbativity at large
$Q^{2}$. Clearly, this condition is qualitative and may be formulated in a
number of ways. We note that the recent lattice calculations of the
running strong coupling constant (see e.g.\ Ref.~\cite{alphaSlattice})
indicate the agreement with perturbative values at the scales of order
5~GeV and higher. We therefore consider, rather arbitrarily, a condition
$M(Q^{2}=25~\mbox{GeV}^{2})\lesssim 0.01$~GeV. Neither the precise form
nor the very existence of this constraint affect the principal
results of this work. This constraint is also shown as a contour
on the $(\mu, \lambda)$ plane in Fig.~\ref{fig:chi2}. The allowed range of
$M(Q^{2})$ is presented in Fig.~\ref{fig:run-mass}
together with some
examples of the allowed functions.

\subsection{Results and discussion}
\label{sec:formfactor:results}

\begin{figure}
\centering
\includegraphics[width=0.95\columnwidth]{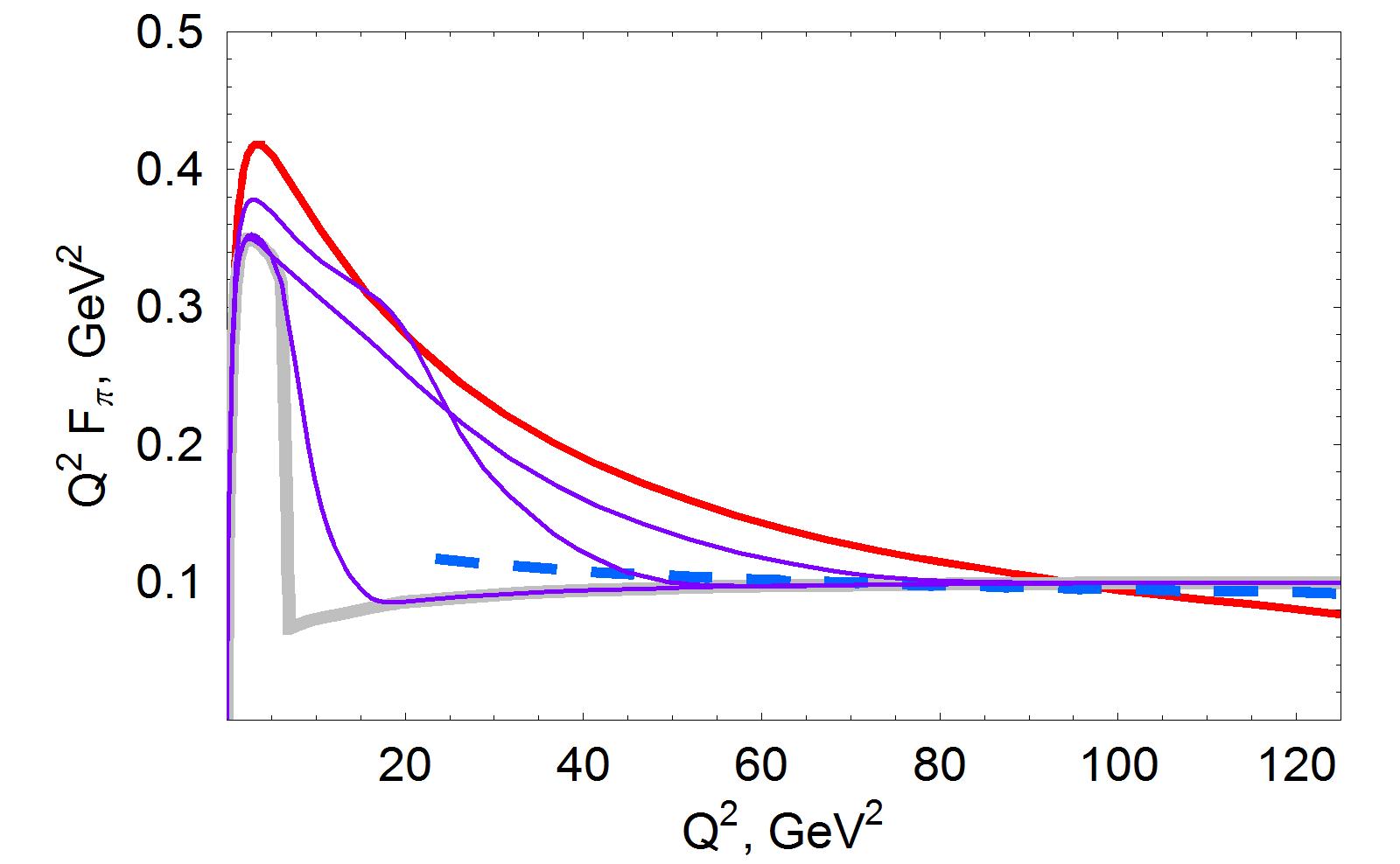}
\caption{Examples of the allowed solutions for $F_{\pi}(Q^{2})$ (thin
lines) demonstrating how the QCD asymptotics, Eq.(\ref{Eq:as}) (dashed
line) settles down. The thick gray line bounds from below the range of all
solutions allowed by the experimental constraints. The thick full (red) line
represents the solution with $M=$const, Refs.~\cite{KrTr-EurPhysJ2001,
KrTr-PRC2009}.}
\label{fig:Fpi125GeV}
\end{figure}

The results of the calculation of $F_{\pi}$ are presented in
Fig.~\ref{fig:Fpi125GeV}.
The freedom in the shape of the transition from $M_{\rm ir}$ to $M_{\rm
uv}$ is illustrated by presenting a few example functions
$F_{\pi}(Q^{2})$ corresponding to the $M(Q^{2})$ functions from
Fig.~\ref{fig:run-mass} which satisfy our constraints.
The lower bound on the allowed $F_{\pi}(Q^{2})$ is determined by
the 95\% C.L. limit from the present experimental data.

The first and immediate conclusion is that the asymptotics~(\ref{Eq:as})
does settle down at $Q^{2}\to\infty$, where $M(Q^{2})\approx 0$. Not only
the correct behaviour
$Q^{2}F_{\pi}(Q^{2}) \simeq \mbox{const}$
is observed (it was expected from analytical calculations of
Ref.~\cite{TMF2005}), but also the coefficient in Eq.~(\ref{Eq:as}) is
reproduced numerically, independently of the selected shape of the
$M(Q^{2})$ function. This, seemingly miraculous, result indicates a deep
connection between low- and high-energy degrees of freedom in our model
which will be discussed elsewhere (see however Sec.~III~G of
Ref.~\cite{long}, an extended version of Ref.~\cite{KrTr-PRC2002}).

The second important implication of our results is that the present
measurements of $F_{\pi}$ constrain the evolution of $M(Q^{2})$ far beyond
the momentum transfers covered by the data. We see that the effective
constituent-quark mass should remain large, $M\sim \Lambda_{\rm QCD}$, at
least up to $Q^{2}\sim 7~\mbox{GeV}^{2}\gg \Lambda_{\rm QCD}^{2}$.
Therefore, nonperturbative dynamics is important at these scales and the
QCD asymptotics of the pion form factor shall start settling down at
even higher values of the momentum transfer.

Let us briefly compare our results with other attempts to describe the
transition from soft to hard behaviour of the pion form factor.

A widely used approach to the calculation of the pion form factor, as well
as of other observables, is based on the light-front quantization, see
Ref.~\cite{LepageBrodskyRev}. In particular, it has been applied to the
calculation of both soft and hard contributions to the pion form factor
more than two decades ago \cite{added}. It was pursued also in
Ref.~\cite{Kiss}, the concept of which is the most close to
ours. The authors of Ref.~\cite{Kiss}, however, did not obtain the
asymptotics (\ref{Eq:as}). This fact may be related to properties of the
constituent-quark model they used: crucial properties (ii) and (iii) of
our model, see Introduction, may not hold for other approaches. An
alternative development of the light-front approach is related to
holography~\cite{BrodskyFpi}. The results of Ref.~\cite{Kiss} illustrate
general trends observed in numerous papers which we cannot review here
(see e.g.\ the review in Ref.~\cite{exp-data} and references in
\cite{BrodskyFpi, 1102.3122}): (1)~the asymptotics $Q^{2}F_{\pi}\sim$const
is not observed and (2)~it is difficult to obtain the decrease of
$Q^{2}F_{\pi}$ down to the QCD values and to satisfy the experimental
constraints simultaneously. A number of successful approaches therefore
give a good description of the data but do not address the high-$Q^{2}$
behaviour of $F_{\pi}$ at all.

\begin{figure}
\centering
\includegraphics[width=0.95\columnwidth]{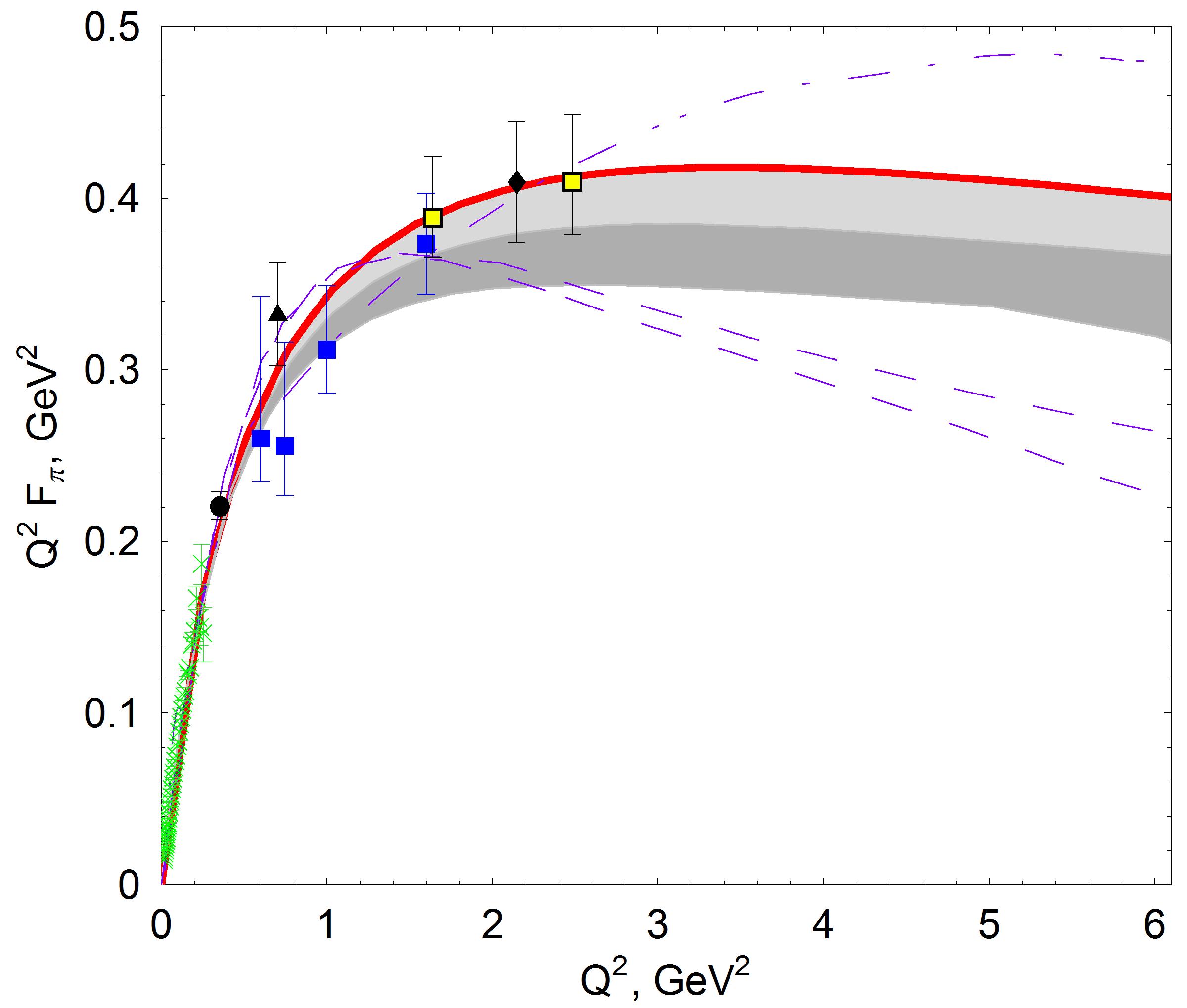}
\caption{
Predictions of this work for $F_{\pi}(Q^{2})$ (gray shade; cf.\
Fig.~\ref{fig:run-mass}) together with $M=$const prediction of the same
model~\cite{KrTr-EurPhysJ2001, KrTr-PRC2009} (full line), and those of
Refs.~\cite{MarisRoberts} (area between two dashed lines) and
Ref.~\cite{newTandy} (dash-dotted line). The experimental data points
from Refs.~\cite{Amendolia} (crosses), \cite{Ackerman} (reanalized in
Ref.~\cite{exp-data}, circles),
\cite{Brauel} (reanalized in Ref.~\cite{exp-data}, triangles),
\cite{Horn} (diamond) and \cite{exp-data}  (squares) are also shown.
}
\label{fig:Fpi6GeV}
\end{figure}

The picture similar to what we observe, that is the $Q^{2}F_{\pi}
\sim$const behaviour at large $Q^{2}$, has been observed only in
Ref.~\cite{MarisRoberts}, cf.\ their Fig.~2. The value of the constant
obtained in that work may be tuned by changing model parameters within
the allowed values. The predictions of Ref.~\cite{MarisRoberts} are shown
in Fig.~\ref{fig:Fpi6GeV} together with our results and the experimental
data.

Another prediction, which we also plot in Fig.~\ref{fig:Fpi6GeV}, has been
obtained in a recent paper~\cite{newTandy}. Instead of attempting to reach
the asymptotics (\ref{Eq:as}), they choose to calculate corrections to it,
which are expected to be large at intermediate values of $Q^{2}$. They
note that these corrections (see e.g.\ Ref.~\cite{LepageBrodskyRev}) are
related to the pion valence-quark distribution amplitudes $\phi_{\pi}(x)$
which differ, at intermediate $Q^{2}$, from their asymptotical values
assumed in Eq.~(\ref{Eq:as}). They use the non-perturbative information
\cite{1306.2645, PRL110-132001} about $\phi_{\pi}(x)$ to obtain the
corrected asymptotics and predict $F_{\pi}(Q^{2})$ in a reasonable
agreement with it at $Q^{2} \gtrsim 10$~GeV$^{2}$. Their approach is
complementary to ours: while we start from a constituent-quark model and
encode nonperturbative dynamics in $M(Q^{2})$, they start from QCD and
encode nonperturbative dynamics in $\phi_{\pi}(x)$. Note that the
perturbative QCD asymptotics, Eq.~(\ref{Eq:as}), is not expected to settle
down before $Q^{2}>1000$~GeV$^{2}$ in their model~\cite{newTandy}.

From Fig.~\ref{fig:Fpi6GeV}, one can see that different approaches give
different predictions for $F_{\pi}(Q^{2})$ already at $Q^{2}\sim
5$~GeV$^{2}$. The expected Jlab upgrade \cite{JLABupgrade} would make it
possible to measure the pion form factor at $Q^{2}\le 6$~GeV$^{2}$ with
the expected precision sufficient to distinguish between the models shown
in Fig.~\ref{fig:Fpi6GeV}.

\section{Conlusions}
\label{sec:concl}
This work addresses the pion form factor at intermediate and large momentum
transfer. The main result of the paper is twofold.

Firstly, we presented a calculation of $F_{\pi}(Q^{2})$ for intermediate
values of $Q^{2}$ and described quantitatively a
range of possible scenarios of the soft-hard transition.
The calculation has been done in the frameworks of a relativistic
constituent-quark model, which had been very successful in predicting the
pion form factor behaviour subsequently measured by Jlab. The model was
supplemented by a model of effective $Q^{2}$-dependent quark mass
motivated by other studies. As a result, we obtained a quantitative
description of the transition from the soft constituent-quark regime to
the hard QCD asymptotics for the pion form factor.
When the quark mass is switched off, the model reproduces the QCD
asymptotics (both the $1/Q^{2}$ behaviour and the coefficient) of the pion
form factor without tuning of parameters, provided the low-energy data are
well described.

Secondly, we demonstrated that the existing $F_{\pi}$ data indicates that
the perturbative QCD fails to describe the pion at momentum transfers at
least as large as $Q^{2} \lesssim 7$~GeV$^{2}$. In particular, the
effective constituent-quark mass does not reach the QCD current-quark
running mass at these values of the momentum transfer.

\begin{acknowledgments}
We are indebted to Stanley Brodsky and Dmitry Levkov for interesting
discussions and comments on the draft. ST thanks CERN (PH-TH division) for
hospitality at the final stages of this work. The work of ST was supported
in part by RFBR (grants 11-02-01528, 12-02-01203, 13-02-01311 and
13-02-01293),  the RF President (grant NS-5590.2012.2) and the RF Ministry
of Science and Education (agreements 8525 and 14.B37.21.0457).
\end{acknowledgments}

\appendix*
\section{Formulae for the form-factor calculation}
The free two-particle form factor  $g_0(s,Q^2,s')$, which enters
Eq.~(\ref{Eq:Fpi-integral}), has the form
$$
g_0(s,Q^2,s')=
  \frac{(s+s'+Q^2)Q^2}{2\sqrt{(s-4M^2)
(s'-4M^2)}}\;
$$
$$
\times \frac{\theta(s,Q^2,s')}{{[\lambda_1(s,-Q^2,s')]}^{3/2}}
\frac{1}{\sqrt{1+Q^2/4M^2}}
$$
$$
\times\left\{(s+s'+Q^2)[G^q_E(Q^2)+G^{\bar q} _E(Q^2)]\right.
$$
$$
\times\cos{(\omega_1+\omega_2)} + \frac{1}{M}\,\xi(s,Q^2,s')
(G^q_M(Q^2)
$$
$$
\left.+ G^{\bar q}_M(Q^2))\sin(\omega_1+\omega_2)\right\}\; ,
$$
where the notations $\lambda_{1}(a,b,c)=a^2+b^2+c^2-2(ab+ac+bc)$ and
$$
\xi=\sqrt{ss'Q^2-M^2\lambda_1(s,-Q^2,s')}\;,
$$
are introduced;
$\omega_1$ and
$\omega_2$ are the Wigner rotation parameters,
$$
\omega_1\!=\!\arctan{\frac{\xi(s,Q^2,s')}{M[(\sqrt s\! +\! \sqrt
{s'})^2\! +\! Q^2]\! +\! \sqrt{ss'}(\sqrt s\! +\! \sqrt{s'})}},
$$
$$
\omega_2\! =\! \arctan{\frac{\alpha(s,s')\xi(s,Q^2,s')}{M(s\!
+\! s'\! +\! Q^2)\alpha( s,s')\! +\! \sqrt{ss'}( 4M^2\! +\!
Q^2)}},
$$
$\alpha(s,s') =  2M+\sqrt s+\sqrt {s'}$, $\theta(s,Q^2,s')=
\vartheta(s'-s_1) -\vartheta(s'-s_2)$, $\vartheta$ is the step
function,
$$
s_{1,2}=2M^2+\frac{1}{2M^2} (2M^2+Q^2)(s-2M^2)
$$
$$
\mp
\frac{1}{2M^2} \sqrt{Q^2(Q^2+4M^2)s(s-4M^2)}.
$$
The functions $G^{u,\bar{d}}_{E,M}(Q^2)$ are the electric and magnetic form
factors of quarks, respectively:
$$
G^q_E(Q^2)=|e_q|f_q(Q^2),
$$
\[
G^q_M(Q^2)=(|e_q|+\kappa_q)f_q(Q^2),
\]
where $q$ denotes the $u$ and $\bar d$ quarks, $e_q$ are their charges,
 $\kappa_q$ are quark anomalous magnetic moments (that, in the end,
 enter our calculation through their sum $s_{q}=\kappa_u+\kappa_{\bar
 d}\approx 0.0268$);
$$
f_q(Q^2)=\frac{1}{1+\ln(1+\langle r_q^2\rangle Q^2/6)}\;,
$$
and the quark mean-square radius $\langle r^2_q\rangle \approx c/M^{2}$
with $c=0.3$ .

\end{document}